\documentclass[twocolumn,a4paper]{article}
\usepackage{epsfig}
\begin{document}
\twocolumn[%
\begin{center}
\textbf{\Large %
Andreev Spectroscopy for Superconducting Phase Qubits
}\vskip1ex
{
M.V.\ Feigel'man$^1$,
V.B.\ Geshkenbein$^{1,2}$,
L.B.\ Ioffe$^{1,3}$, and
G.\ Blatter$^2$
}
\vskip0.5ex
\textsl{ %
$^1$ L. D. Landau Institute for Theoretical Physics, Moscow 117940, Russia\\
$^2$ Institute fur Theoretische Physik, ETH-H\"onggerberg, CH-8093,
Switzerland\\
$^3$ Department of Physics, Rutgers University, Piscataway,
NJ 08855, USA
}
\end{center}
]
Solid state implementations of qubits are challenging, as macroscopic
devices involve a large number of degrees of freedom and thus are
difficult to maintain in a coherent state. This problem is less accute
in designs based on superconducting (SC)\ electronics, which can be
divided into two broad classes: the ``charge'' qubits encode different
states through the charge trapped on a SC island
\cite{Makhlin99,Nakamura99}, while in the ``phase'' qubits the states
differ \emph{mostly} by the value of the phase $\varphi$ of a
superconducting island in a low-inductance SQUID
loop~\cite{Ivanov98,Ioffe99}. As a consequence of long-range Coulomb 
forces, the charge qubit interacts strongly with the environment 
and with other qubits. On the contrary, the phase qubit is more 
effectively decoupled from the environment (a \emph{pure} 
phase qubit with states differing \emph{only} by the value 
of $\varphi$ has practically zero interaction with the environment). 
Although pure phase qubits can be fabricated
from $d$-wave superconductors~\cite{Ioffe99,Ilichev98} the task is
technologically demanding.  Here, we concentrate on the most simple
version of a superconducting phase qubit (SPQB), a Josepshon loop made
from a few submicron sized SC islands connected via similar Josephson
junctions and placed in a frustrating magnetic field; such a device
with 4 junctions is sketched in Fig.\ 1.  Two degenerate states
naturally appear in such a loop if the flux $\Phi _{q}$ of the
external field through the qubit loop is exactly $\Phi _{0}/2=hc/4e$.
Using a gauge $A_{x}=0$, $A_{y}=H x$, the classical minima of the
Josephson energy are attained when the phase on the island $G$
(Fig.~1) takes the value $\varphi ^{\pm }=\pm \pi /2$ (relative to the
phase at the point $O$); below we refer to these states as
$|\!\!\uparrow \rangle$ and $|\!\!\downarrow \rangle$.
\begin{figure}[hbt]
\centering
\epsfig{file=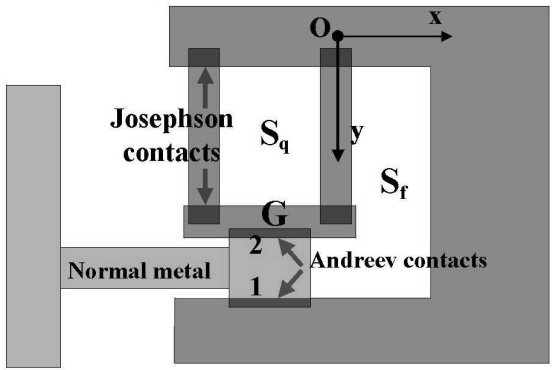,width=\columnwidth}
\caption{SPQB device with Andreev probe; the area $S_q$ of the
``intrinsic'' loop of the qubit is equal to the area $S_{f}$ of the
NS-QUID loop.}
\label{fig1}
\end{figure}
In order to reduce any parasitic coupling to the environment, the
inductance $L$ of the loop shall be small with $LI_{c}\leq
10^{-3}\Phi_{0}$.  Ignoring charging effects, the system prepared in
one of these states will stay there forever; quantum effects appear
when the charging energies are accounted for. They are determined by
the capacitances $C$ of the junctions and we require them to be
smaller than the Josephson energy, $e^{2}/C \ll \hbar I_{c}/e$. The
tunnelling rate between the two classically degenerate ground
states is estimated as $\Omega \approx \sqrt{eI_{c}/\hbar
  C}\exp(-a\sqrt{\hbar I_{c}C /e^{3}})$, where $a$ is of order 1. We
assume values of $I_{c}$ between $10-100$ nA and capacitances of order
of a few fF, resulting in a characteristic Josephson plasma frequency
$\omega_{pl}\sim 100$ GHz and a tunneling frequency $\Omega \sim 1-10$
GHz. Once tunnelling is taken into account, the true eigenstates
become $|0\rangle=\frac{1}{\sqrt{2}}(|\!\uparrow \rangle+|\!\downarrow
\rangle)$ and $|1\rangle=\frac{1}{\sqrt{2} }(|\!\uparrow
\rangle-|\!\downarrow \rangle)$, separated by an energy gap $\hbar
\Omega$. A deviation of the magnetic flux $\Phi_{q}$ through the loop
from a value $\Phi _{0}/2$ removes the degeneracy of the states
$|\!\uparrow \rangle$ and $|\!\downarrow \rangle$.  The Hamiltonian of
our qubit written in the basis $|\!\uparrow \rangle, |\!\downarrow
\rangle$ takes the form $H=h_{x}\sigma_{x}+h_{z}\sigma_{z}$, where
$h_{z}=\frac{\hbar I_{c}}{2e}(2\Phi_{q}-\Phi_{0})$ and
$h_{x}=\hbar\Omega$.  Varying the effective fields $h_{z}$ and $h_{x}$
we can perform all necessary operations on the qubit. Changing the
external flux through the loop produces a variation in $h_{z}$. The
coupling parameter $h_{x}$ can be smoothly modified through a variation 
of the gate potential applied to the island $G$ (cf.~\cite{Ivanov98}).
Alternatively, short circuiting a junction with an external capacitor
$C_{ext}\sim 10$ $C$ leads to an abrupt blocking of the tunneling
channel (switching off $h_{x}$).

The first task to be addressed in the study of a SPQB is the
development of a convenient probe testing for coherent Rabi
oscillations in the device; this can be accomplished by a measurement
of the phase-sensitive subgap Andreev conductance. The low-$T$
conductance of a NS boundary is determined by the time the incident
electron and the Andreev-reflected hole can interfere
constructively~\cite{Nazarov93,Pothier94}.  The phase coherent
electron diffusion in the normal wire of a ``fork''
geometry~\cite{Nazarov93,Pothier94} leads to periodic conductance
oscillations as a function of the magnetic flux penetrating the region
$S_f$ of the fork (with period $\Phi _{0}$). This is due to the
magnetic field controlling the superconducting phase difference
between the two NS contacts of the fork, thus influencing the electron
interference pattern.

The experiment discussed here is governed by similar physics.
Consider first the subgap conductance between the active SC island of
the qubit and the dirty normal metal wire, connected to the island via
a high-resistance tunnel barrier with normal state conductance
$\sigma_T \leq e^2/\hbar$. If the qubit phase $\varphi$ does not
fluctuate in time (i.e., $\Phi_{q} \neq \Phi_0/2$, $h_z \gg h_x$), the
Andreev conductance at $T \to 0$, $V \to 0$ is $\sigma_A^{cl} =
\sigma_T^2R_D$, where $R_{D} = \rho L \ll \sigma_T^{-1}$ is the
resistance of a dirty wire of length $L$. At voltages $eV \geq \hbar
D/L^2$ the differential subgap conductance $dI_A/dV =\sigma_A^{cl}(V)
\sim \sigma_T^2 \widehat{C}(2eV)$, where $\widehat{C}(E)$ is the
space-integrated Cooperon amplitude~\cite{Nazarov93}.
In a single-wire geometry and in
the absence of decoherence inside the normal wire (i.e., at $T=0$),
the Cooperon amplitude $\widehat{C}(E) = \rho \sqrt{\hbar D/E}$ at $E
\gg \hbar D/L^2$.  For the qubit tuned to resonance, $h_z \leq h_x$,
the time-dependent fluctuations of the SC order parameter destroy the
coherence between multiple Andreev reflections at the NS boundary,
thereby suppressing the subgap conductance. Quantitatively, their
effect on the subgap conductance $\sigma_A(V)$ is
\begin{equation}
\frac{\sigma _{A}(V)}{\sigma _{A}^{cl}(V)}=\frac{\int_{0}^{2eV}dEP(E)%
\widehat{C}(2eV-E)}{\widehat{C}(2eV)}.  \label{1}
\end{equation}
Here, $P(E)=\frac{1}{2\pi}\int e^{iEt}K(t)dt$ and $K(t)=\langle
e^{i(\varphi (0)-\varphi (t))}\rangle $ is the intrinsic correlation
function of the SPQB.  In the case of weak decoherence, $K(t)$ is
given by $K(t)=e^{-i\Omega t}e^{-\Gamma |t|}$, with $\Omega$ the
tunnelling frequency and $\Gamma$ the intrinsic decoherence rate. The
derivation of Eq.~(\ref{1}) parallels the one presented
in~\cite{Huck97,Feigel99}.  In the case of completely coherent Rabi
oscillations ($\Gamma\to 0$), the conductance $\sigma_A(V)$ vanishes at
$2eV < \hbar\Omega$ and exhibits a square-root singularity
$\sigma_A(V) \propto 1/\sqrt{2eV-\hbar\Omega}$ 
above the thres\-hold,
i.e., it behaves as a tunneling conductance into a BSC superconductor
at $2eV > \Delta_{BCS}$. In the opposite limit of incoherent quantum
tunnelling of the phase ($\Gamma \gg \Omega$), the ratio (\ref{1})
decreases as $eV/\hbar\Gamma$ at low voltages.

Next, consider the full interference experiment for the device shown
in Fig.~1. The idea is to measure the amplitude $I_A^{(12)}$ in the
oscillations of the Andreev current $I_A = I_A^{(1)} + I_A^{(2)} +
I_A^{(12)}$ as a function of the magnetic field $H$ (where the
superscripts $^{(i)}$ refer to the contribution from the contacts
$i=1$ and $i=2$, see Fig.~1).  The total phase determining the
interference current $I_A^{(12)}$ is equal to $\phi_A = 2\pi
HS_f/\Phi_0 + \varphi$.  Away from the degeneracy field $H_n = 
(n+1/2)\Phi_0/S_q$, the phase $\varphi$ in our device is 
determined by the minimization of the Josephson energy and we 
obtain $\varphi = \pi ( \{H S_{q}/\Phi_0 + 1/2\} -1/2)$, 
where $\{x\}$ is the fractional part of $x$.  Choosing a
geometry with $S_q=S_f$, we find for $H S_{q}/\Phi_0 \in
[n,n+1/2]$ an interference current $I_A^{(12)}= j_{12}\cos(3\pi H
S_q/\Phi_0)$, whereas for $H S_{q}/\Phi_0 \in [n+1/2,n+1]$ the sign
changes: $I_A^{(12)}=-j_{12}\cos(3\pi H S_q/\Phi_0)$. Thus, the whole
semiclassical $I_A^{(12)}(H)$ dependence has a period $\Phi_0/S_q$,
with upward cusps at $H = H_n$. These cusps are due to the 
penetration of a flux quantum $\Phi_0$ into the qubit loop 
at the degeneracy points $H_n$. At these fields the interference 
contribution $I_A^{(12)}$ vanishes (in the geometry with $S_q=S_f$) 
and the semiclassical Andreev current is
given by the sum $I_A^{(1)} + I_A^{(2)}$. However, at $H = H_n$ the
quantum fluctuations in the phase $\varphi$ become large and suppress
the Andreev current $I_A^{(2)}$ to the ``active" island G of the SPQB.
This implies that close to the degeneracy points $|H-H_n| \leq
H_0\hbar\Omega/E_J$ a very narrow dip is superimposed upon the
above-mentioned cusp in the $I_A(H)$ dependence.  The minimum value of
the current in this dip coincides with the current $I_A^{(1)}$ to the
``ground''.  The voltage dependence of the dip $dI_A/dV$ can be
related, via Eq.~(\ref{1}), to the autocorrelation function $K(t)$ of
the SPQB.

The proposed measurement is ``non invasive'', i.e., it does not, by
itself, destroy the coherent dynamics of the phase $\varphi$. This is
due to the fact that at low $V$ this experiment does not measure the
value of $\varphi$. In order not to destroy the quantum tunneling of
$\varphi$ during the measurement, the subgap conductance should be
low, i.e. $\sigma_{A}^{-1}\gg 4e^{2}/h=6.5$ k$\Omega$. However, it
should not be too low to be measurable with good accuracy at low
temperatures (i.e., $\sigma _{A}^{-1}\leq 10$ M$\Omega$).
Well-controlled metallic thin films with a long decoherence time can
be produced with a sheet resistance $ \sim 0.1$ k$\Omega$; the use of
such films require $ \sigma _{T}^{-1}$ in the range $3-30$ k$\Omega$
in order to place $\sigma _{A}^{-1}$ within a range $0.1-10$
M$\Omega$.



\end{document}